\documentstyle[aps,twocolumn]{revtex}

\title{Iterative algorithm for  reconstruction of entangled states}
\author{J. \v{R}eh\'{a}\v{c}ek\cite{email}, Z. Hradil, and M. Je\v{z}ek}
\address{Department of Optics, Palack\'{y} University, 17. listopadu 50,
772 00 Olomouc, Czech Republic}

\begin{document}

\maketitle

\begin{abstract}
An iterative algorithm for the reconstruction of an unknown quantum
state from the results of incompatible measurements is proposed. It
consists of Expectation-Maximization step followed by a  unitary
transformation of the eigenbasis of the density matrix. The procedure
has been applied to the reconstruction of the entangled pair of
photons.

\end{abstract}

\section{Introduction}

Predictions of advanced theories are more and more  complex and
more and more  accurate. This may be recognized in recent
progress of  quantum theory. In many applications there is a need
to determine the quantum state of the system. For this purpose the
quantum tomography has been developed.  There is an extended
bibliography concerning this topic  covering various fields of
possible applications \cite{prehled}. However, all the varied and
precise experiments which have been carried out over many years and
which rely on quantum physics have not need the total amount of
information coded in quantum state. This fact is reflected in the
standard treatment of quantum tomography. When the standard quantum
tomography is adopted for reconstruction of the state from
realistic noisy data, one often runs into unphysical results.
Standard methods are also known to be prone to creation of various
artifacts in the reconstructed state. All these flaws are usually
paid little or no attention in the scientific literature. They are
simply being regarded as unavoidable errors of reconstructions,
which fall within the corresponding ``error bars''. Here we would
like to stress that the mentioned drawbacks of the standard
tomographic techniques are actually much more serious, especially
if the reconstructed state is to be of further use.
 There might be no clue as to which modifications to the
reconstructed ``state'' should be done in order to make its density
operator semi-positive definite and retain the efficiency of the
reconstruction as high as possible at the same time. This seems to
be crucial for the potential application in quantum information. To
quantify  a fragile effect of entanglement, various entropic
principles are used \cite{entangled}. This feature is sensitive to
semi-positive definiteness of the reconstructed state - a necessary
condition of successful reconstruction.

The purpose of this Rapid Communication is twofold. At first a
simple iterative algorithm for maximum likelihood (ML) estimation
of quantum state resembling  ``climbing the hill of the
likelihood" will be derived. This result may be easily implemented
 numerically, and  interpreted in quantum theory as generalized
measurement.  The algorithm will be illustrated on the example of
reconstruction of entangled state, representing an important
example in quantum information processing.

Let us illustrate the motivation of quantum tomography  considering
the repeated measurement. Assume  that we are given a finite number
$N$ of identical samples of the system, each in the same but unknown
quantum state described by
the density operator $\rho$. Given those systems our task
is to identify the unknown {\em true} state $\rho$ from the results
of measurements performed on them as accurate as possible.
For simplicity we will assume
sharp measurements in the sense of von Neumann.
As a result of each measurement the state of the input system is
projected into a pure state, which is the reading of the measuring
apparatus. Let us assume, for concreteness,
that $M$ different outcomes of measurements have been observed.
The relative frequencies
$f_j$ of occurrences of the observed results
\begin{equation}  \label{projekce}
\{|y_j\rangle\langle y_j|\}, \qquad j=1,\dots,M,
\end{equation}
then comprise the data which the true state $\rho$ should be
inferred from. For the sake of simplicity, the measurement
performed will be assumed as complete, i.e.
$$  H \equiv  \sum_j \{|y_j\rangle\langle y_j|\} =  1. $$
This condition will be released later to the case of incomplete
measurements.

The probabilities of occurrences of various outcomes
are generated by the true quantum state $\rho$ according to
the well-known handy quantum rule
\begin{equation} \label{teorie}
p_j=\langle y_j|\rho|y_j\rangle.
\end{equation}
If the probabilities $p_j$ of getting a sufficient number of
different outcomes $|y_j\rangle$ were known, it would be possible
to determine the true state $\rho$  directly by inverting the
linear relation (\ref{teorie}). This the philosophy behind the
``standard'' quantum tomographic techniques. For example, in the
rather trivial case of a spin half particle, the probabilities of
getting three linearly independent projectors determine the
unknown state uniquely. Here, however, a serious problem arises.
Since  only a finite number of systems can be investigated, there
is no way how to find out these probabilities. The only data one
has at disposal are the relative frequencies $f_j$, which sample
the {\em principally} unknowable probabilities $p_j$. It is obvious
that for small number of runs the true probabilities $p_j$ and the
corresponding detected frequencies $f_j$ may differ substantially.
As a result of this, the modified realistic problem
\begin{equation}    \label{problem}
f_j=\langle y_j|\rho|y_j\rangle
\end{equation}
has generally no solution on the space of semi-positive definite
hermitian operators describing physical states.

Probabilistic interpretation of quantum theory suggests that a sort
of statistical treatment of the observed data might be more natural
and appropriate than the deterministic treatment described above.
The philosophy of the reconstruction method differs from the
philosophy of standard methods. The basic question of the standard
methods: ``What quantum state is determined by the measured data?''
is replaced with a more modest one: ``What quantum state is most
likely in view of the measured data?''; this seems to be in
accordance with the probabilistic interpretation of the quantum
theory \cite{hradil}. More specifically, instead of trying to
invert the linear relation (\ref{problem}), we look for a density
operator $\rho_e$, which generates through Eq.~(\ref{teorie})
probabilities $p_j$ that are as ``close'' to the observed
frequencies $f_j$ as possible, i.e.
\begin{equation}  \label{reseni}
\rho_e=\arg\{\min\limits_{\rho} d[{\bf f},{\bf p}(\rho)]\},
\end{equation}
where $d[.,.]$ stands for some measure of distance between the two
probability distributions ${\bf p}$ and data ${\bf f}$.
 At first sight it might seem that there is no reason to {\em
prefer} one particular metric to another one -- different metrics
leading to different results through Eq.~(\ref{reseni}). This
ambiguity can be resolved by considering the formal description of
the reconstruction process \cite{zdenek_fund}. If the whole
measurement and subsequent reconstruction is looked at as a single
generalized measurement, then the {\em relation} between the
actually performed measurement and resulting probability operator
measure becomes particularly simple and easy to interpret for the
metric known as the relative entropy or Kullback-Leibler divergence
\cite{Kullback}
:
\begin{equation} \label{leibler}
d[{\bf f},{\bf p}]=-\sum_j f_j \ln p_j.
\end{equation}
Solving the problem (\ref{reseni}) with the metric (\ref{leibler})
is equivalent to finding the maxima of the likelihood functional
\begin{equation} \label{ml}
{\cal L}(\rho)=\prod_j \langle y_j|\rho|y_j\rangle^{f_j}.
\end{equation}
Thus we are led to the Maximum Likelihood  principle as the
preferred way of doing the quantum state reconstruction.

 ML methods are well-known in the field of inverse problems and
they have found many applications in reconstructions and
estimations so far \cite{ideal}. Unfortunately, except in most
simple cases, the maximization of the likelihood functional is a
challenging problem on its own. A necessary condition for an
extreme of the likelihood functional (\ref{ml}) can be derived in
the form of the nonlinear operator equation for the density matrix
$\rho$ \cite{hradil,zdenek_nelin} and this equation may be
interpreted as closure relation for a quantum measurement. In the
classical signal processing an important role is played by the so
called linear and positive (LinPos) problems \cite{em}. Since these
are closely related to the problem of quantum state reconstruction
it is worthwhile to recall how the positive and linear problems can
be dealt with using the ML approach.

Let us consider that the probabilities $p_j$ of getting outcomes $y_j$
are given by the following linear and positive relation
\begin{equation} \label{linpos}
p_j=\sum_i r_i h_{ij}, \quad \bf{p},\bf{r},\bf{h}> 0.
\end{equation}
Here ${\bf r}$ is the vector describing the ``state'' of the
system. For example, the reconstruction of a one-dimensional object
from the noiseless detection of its  blurred image could be
accomplished by inverting the relation (\ref{linpos}), where  ${\bf
r}$ and ${\bf p}$ would be the normalized intensities of the object
and image, and ${\bf h}$ would describe the blurring mechanism.
Again here the presence of noise ($f_j\neq p_j$) tends to spoil the
positivity of the reconstructed intensity ${\bf r}$.

The solution to LinPos problems in the sense of
(\ref{reseni}) can be found using the Expectation-Maximization (EM)
algorithm \cite{em}. In the case of the discrete one-dimensional
problem (\ref{linpos}) the unknown object ${\bf r}$ is
reconstructed by means of the following iterative algorithm
\cite{em}:
\begin{equation} \label{iter}
r_i^{(n)}=r_i^{(n-1)} \sum_j \frac{h_{ij} f_j}
{p_j({\bf r}^{(n-1)})},
\end{equation}
which is initialized with a positive vector ${\bf r}$ ($r_i
> 0\, \forall i$).

The iterative algorithm (\ref{iter}) for solving the LinPos
problems is convenient from the point of view of the numerical
analysis. It is certainly much more convenient than the direct
multidimensional maximization of the corresponding ML functional
$\ln{\cal L}=\sum_j f_j \ln p_j$ \cite{banaszek}. This brings us
back to the problem of quantum state reconstruction. It would be
nice to have a similar iterative algorithm for dealing with the
problem (\ref{problem}) [\,or equivalently for maximizing the ML
functional (\ref{ml})]. On the one hand it is clear that the
problem of quantum state reconstruction is not a linear
{\em and} positive problem, since the quantum rule (\ref{teorie})
cannot be rewritten to the form of Eq.~(\ref{linpos})
with a known positive kernel ${\bf h}$.
As a consequence of this the
EM algorithm cannot be straightforwardly applied here. On the other
hand the reconstruction of the elements of the density matrix
becomes a LinPos problem if the eigenbasis diagonalizing
the density matrix is  known. In this case the unknown density
matrix can be parametrized as follows
\begin{equation} \label{fixrho}
\rho=\sum_k r_k |\phi_k\rangle\langle\phi_k|,
\qquad \rho|\phi_k\rangle=r_k|\phi_k\rangle,
\end{equation}
where $r_i$ are eigenvalues of $\rho$, the only parameters which
remain to be determined from the performed measurement. Using the
parameterization (\ref{fixrho}) the quantum rule (\ref{teorie}) may
be easily rewritten to the form of LinPos problem
Eq.~(\ref{linpos}).

This hints on splitting the quantum state reconstruction into two
subsequent steps: the reconstruction of the eigenvectors of $\rho$
in a fixed basis, which represents the classical part of the
problem, followed by the ``rotation'' of the basis
$\{|\phi_i\rangle\}$ in the  ``right'' direction using the unitary
transformation
\begin{equation} \label{novephi}
|\phi_k'\rangle\langle\phi_k'|=U|\phi_k\rangle\langle\phi_k|U^{\dag}.
\end{equation}
Its infinitesimal form reads
\begin{equation} \label{transform}
U\equiv e^{i\epsilon G}\approx 1+i\epsilon G.
\end{equation}
Here $G$ is a Hermitian generator of the unitary transformation
and $\epsilon$ is a positive real number which is
small enough in order to make the second equality
in (\ref{transform}) approximately satisfied.

Consider now the total change of log-likelihood caused by the
change of diagonal elements of density matrix and rotation of
basis. Keeping the normalization condition  $Tr  \rho = 1$, the
first order contribution to the variation reads
\begin{eqnarray} \label{lnl}
\delta \ln{\cal L}(\rho')&=& \sum_k \delta r_k
(\langle \phi_k|  R |\phi_k \rangle -\lambda)\nonumber \\ &&\qquad
+ i\epsilon {\rm
\,Tr}\left\{G[\rho,R]\right\}.
\end{eqnarray}
The operator $R$ appearing here plays an important role in this
treatment. It is semi-positively definite Hermitian operator
comprising results of the measurement
\begin{equation} \label{rozklad}
R=\sum\limits_j \frac{f_j}{p_j}|y_j\rangle\langle y_j|.
\end{equation}
Notice this operator depends on the old density matrix
$\rho$ through Eq.~(\ref{teorie}).

Inspection of Eq.~(\ref{lnl}) reveals a simple strategy how to make
the likelihood of the new state $\rho'$ as high as possible [within
limits of the validity of the linearization (\ref{transform}), of
course]. In the first step the first term on the right hand side of
Eq.~(\ref{lnl}) is maximized by estimating the eigenvalues of the
density matrix keeping its eigenvectors $|\phi_k\rangle$ constant.
The iterative algorithm (\ref{iter}) described above  can
straightforwardly be applied to this LinPos problem.  In the second
step the likelihood can further be increased by making the second
term on the right hand side of Eq.~(\ref{lnl}) positive. This is
accomplished by a suitable choice of the generator of the unitary
transformation (\ref{transform}). Reminding the norm induced by the
scalar product defined on the space of operators , $(A,B)={\rm
Tr}\{A^{\dag}B\}$, the generator $G$ may be  chosen as
\begin{equation} \label{komut}
G=i[\rho, R].
\end{equation}
 Its form  guarantees the  non-negativity of the contribution to
the likelihood.  This choice is also optimal in the sense of the
above introduced scalar product.

Now we have at our disposal all ingredients comprising the EMU
quantum state reconstruction algorithm which represents the main
result of the present article. Starting from some positive initial
density matrix $\rho$ this estimate is improved, first by finding
new eigenvalues using the EM iterative algorithm (\ref{iter}), and
then again by finding new eigenvectors by unitary (U) transforming
the old ones according to Eqs.~(\ref{novephi}-\ref{transform}) and
(\ref{komut}). These two steps are repeated. Continued repetition
of the two steps, each monotonically increasing the likelihood of
the current estimate, resembles of climbing a hill.

The proposed EMU algorithm naturally leads to the previously
introduced extremal equation for the  density matrix
\cite{hradil,zdenek_nelin}. The stationary  point of  EMU algorithm
is characterized by the vanishing variance of the log likelihood
(\ref{lnl}). Since the variations $\delta r_k, \epsilon$ are
arbitrary parameters, this is equivalent to the Lagrange-Euler
equation for density matrix
\begin{equation} \label{nelin}
R\rho_e=\rho_e.
\end{equation}
This nonlinear operator equation has recently been derived  using
the variational principle in Ref. \cite{zdenek_nelin} and using the
inequalities \cite{hradil}. The EMU algorithm presented here thus
provides us with a different route to its derivation, which is
perhaps more appealing from the physical point of view, and
suitable for implementation of numerical algorithm. Notice,
however, that this ``parameter estimation" may be interpreted as a
generalized measurement since  $ R =  1$ on the space where the
reconstruction has been done.

These results should  be modified in  the case of incomplete
detection. Provided that $ H \neq  1,
$
the closure relation may be always recovered  in the form
\begin{equation}
\sum_j  H^{-1/2}  \{|y_j\rangle\langle y_j|\}  H^{-1/2} \equiv
1.
\end{equation}
This  corresponds to the   renormalization of the probabilities
$p_j
= \langle y_j | \rho |  y_j \rangle$ in the likelihood
(\ref{ml}) to the normalized probabilities
$p_j \rightarrow p_j/\sum_i p_i$.
 This  formulation  incorporates  the case of incomplete detection.
Notice, that extremal equation  possesses again the form of
equation
 (\ref{nelin}) for the renormalized quantities
\begin{eqnarray} \label{renormalization}
 R \rightarrow  R^{\prime} =   (H^{\prime})^{-1/2}  R (
H^{\prime})^{-1/2}, \nonumber\\  \rho_e \rightarrow
\rho_e^{\prime} = ( H^{\prime})^{1/2}
 \rho_e ( H^{\prime})^{1/2}, \nonumber\\  H^{\prime} =
\frac{1}{\sum_j p_j} \sum_j \{|y_j\rangle\langle y_j|\}.
\end{eqnarray}
All the conclusions derived for complete measurements may be
extended to this case of incomplete measurement as well. This
formulation coincides with the estimation provided that assumption
of Poissonian statistics is used. Assume that $n_i$ samples the
mean number of particles $n p_i,$ where $p_i $ is as before the
prediction of quantum theory for detection the $i$-th channel and
$n$ is unknown mean number of particles. The relevant part of
log-likelihood corresponding to the Poissonian statistics reads
\begin{equation}
\ln {\cal L} \propto \sum_i n_i \ln (n p_i) -n \sum_i p_i.
\end{equation}
The extremal equation for n may be easily formulated as the
condition
$n = \sum_i n_i/\sum_i p_i$.
Inserting this estimate of unknown mean number of Poissonian
particles into the log-likelihood reproduces the renormalized
likelihood function.

The proposed EMU algorithm has been applied to the reconstruction
of two-photon entangled state generated by the spontaneous
downconversion source of White et al. \cite{white}. White et al.
measured the nominal state $|HH\rangle+|VV\rangle$ along sixteen
distinct directions: $\{|y_j\rangle\}=\{|HH\rangle$, $|HV\rangle$,
$|VH\rangle$, $|VV\rangle$, $|HD\rangle$, $|HL\rangle$,
$|DH\rangle$, $|RH\rangle$, $|DD\rangle$, $|RD\rangle$,
$|RL\rangle$, $|DR\rangle$, $|DV\rangle$, $|RV\rangle$,
$|VD\rangle$, $|VL\rangle\}$. $H$, $V$, $D$, $R$, and $L$ being
horizontal, vertical, diagonal, right circular, and left circular
polarization, respectively. Counted numbers of coincidences along
these directions can be found in \cite{white}.

We have used the experimental data together with the
proposed algorithm to reconstruct
the {\em true} state generated by the source of entangled photon
pairs.
Due to various sources of errors the true state is expected to
differ from the nominal state.
Notice that the chosen measurements are not complete,
that is $\sum_j|y_j\rangle\langle y_j|$ does not represents
the resolution of unity. This has been taken into account,
see Eq.~(\ref{renormalization}).

Starting from maximally mixed state
$(|HH\rangle\langle HH|$ + $|VV\rangle\langle VV|$ +
$|HV\rangle\langle HV|$
+ $|VH\rangle\langle VH|)/4$,
new eigenvalues and eigenvectors of density matrix are
found using Eqs.~(\ref{iter}) and (\ref{novephi}).
This has been repeated until a stationary point of the iteration
process has been attained.
The diagonal representation of the reconstructed
density matrix reads
\begin{equation} \label{nase}
\rho_e^{\rm ML}=0.962\,|\phi_1\rangle\langle \phi_1|+
0.038\,|\phi_2\rangle\langle \phi_2|.
\end{equation}
The other two eigenvalues are zero.
The eigenvectors $|\phi_1\rangle$ and $|\phi_2\rangle$
are given in Tab.~(\ref{tab}).

The reconstructed density matrix (\ref{nase}) agrees well
with the qualitative reasoning given in \cite{white}.
Namely, the reconstructed state is almost pure state --
slightly rotated nominal state $|HH\rangle+|VV\rangle$.
The apparent incompatibility of the nominal state with
the registered data was interpreted in \cite{white} as the
result of possible slight misalignments of the axes of
analysis systems with respect to the axes of the downconversion
source. This is, of course, reflected in the reconstructed state
(\ref{nase}), which quantifies such misalignments and might
serve for hunting down the errors, and calibrating the experimental
setup.

Notice also that the reconstructed density matrix (\ref{nase})
is semi-positive definite. This should be contrasted
with the result of standard reconstruction.
Direct inversion of Eq.~(\ref{problem}) yields the
density matrix having the following diagonal
representation \cite{private}
\begin{equation} \label{jejich}
r_1=1.022,\,\, r_2=0.068,\,\, r_3=-0.065,\,\, r_4=-0.024
\end{equation}
The corresponding eigenvectors need not be specified here.
Apparently, direct inversion (standard tomography) leads to
unphysical non-positive definite result. It is worth to notice that
the negative eigenvalues are comparable in magnitude with
non-diagonal elements of $\rho_e^{\rm ML}$ in $H$-$V$ basis, see
Tab.~\ref{tab}. This is a nice example of situation when standard
methods fail even though rather high number of particles (tens of
thousands) has been registered. ML reconstruction provides always
physically sound results. Moreover, it represents genuine quantum
measurement of entangled state.

 This work was supported by TMR Network ERB FMRXCT 96-0057
``Perfect Crystal Neutron Optics'' of the European Union and by
projects CEZ J14/98 and LN00A015. This paper is dedicated to the
anniversary of 65th birthday of Prof. Jan Pe\v{r}ina.

\begin{table}
\begin{tabular}{lcc}
& $|\phi_1\rangle$ & $|\phi_2\rangle$ \\
\hline\\
$|VV\rangle$ & $0.696-0.027i$ & $0.630+0.071i$\\
$|VH\rangle$ & $-0.050-0.020i$ & $-0.284+0.174i$ \\
$|HV\rangle$ & $-0.040+0.015i$ & $-0.150-0.247i$\\
$|HH\rangle$ & $0.712-0.062i$ & $-0.634-0.035i$
\end{tabular}
\caption{Eigenvectors of the reconstructed density matrix.}
\label{tab}
\end{table}

\end{document}